\documentclass[aps,prb,twocolumn,groupedaddress]{revtex4}
\usepackage{graphicx}% Include figure files
\usepackage{dcolumn}% Align table columns on decimal point
\usepackage{bm}% bold math

\begin{document}
%\draft
\title{Superconducting state  in the non-centrosymmetric Mg$_{9.3}$Ir$_{19}$B$_{16.7}$ and Mg$_{10.5}$Ir$_{19}$B$_{17.1}$ revealed by NMR}
\author{K. Tahara $^{1}$,  Z. Li $^{1}$, H.X. Yang$^{2}$, J.L. Luo$^{2}$, S. Kawasaki$^{1}$ and Guo-qing Zheng $^{1,2,*}$}
\address{ $^1$ Department of Physics, Okayama University, Okayama 700-8530, Japan}
\address {$^2$ Institute of Physics, Chinese Academy of Sciences, Beijing 100190, China }

\date{\today}
%\twocolumn[

\begin{abstract}
We report $^{11}$B NMR measurements in non-centrosymmetric superconductors Mg$_{9.3}$Ir$_{19}$B$_{16.7}$ ($T_c$=5.8 K) and Mg$_{10.5}$Ir$_{19}$B$_{17.1}$ ($T_c$=4.8 K). The spin lattice relaxation rate and the Knight shift indicate that the Cooper pairs are  predominantly in the spin-singlet state with an isotropic gap. However, Mg$_{10.5}$Ir$_{19}$B$_{17.1}$ is found to have more defects and the   spin susceptibility remains finite even in the zero-temperature limit. We interpret this result as  that the defects  enhance the spin-orbit coupling and bring about more spin-triplet component. 
\end{abstract}
%]
\narrowtext
\maketitle
%\widetext

\section{Introduction}

Superconductors without spatial inversion symmetry in the crystal structure have attracted much attention.
In superconducting materials with an inversion center,   the Cooper pairs are either in the spin-singlet state, or in the spin-triplet state, due to Pauli exclusion principle. 
 However, when the  inversion symmetry is broken, the spin-singlet and spin-triplet states can be mixed \cite{Gorkov,Frigeri,Fr}. This was actually found in Li$_2$Pt$_3$B  by nuclear
magnetic resonance (NMR) (Ref.  \cite{Nishiyama}) and other measurements \cite{Yuan,Takeya}.
The extent of parity mixing depends on the strength of the spin-orbit coupling (SOC) that is enhanced by the lack of inversion symmetry. In such materials, novel superconducting properties can be expected \cite{Yip,Sa}. 

After the discovery of the  non-centrosymmetric compound CePt$_3$Si \cite{Bauer}, many new superconductors of such kind have been discovered. They can be categorized into two types. Namely, the strong-correlated electron systems such as  UIr  \cite{Akazawa}, CeRhSi$_3$ \cite{CeRhSi3}, CeIrSi$_3$ \cite{CeIrSi3}, and  the weakly-correlated electron systems that include Li$_2$Pd$_3$B and Li$_2$Pt$_3$B \cite{Togano1,Togano2}, Mg$_{10}$Ir$_{19}$B$_{16}$ \cite{Klimczuk}, Y(La)$_2$C$_3$ \cite{Amano,Simon}, Rh(Ir)$_2$Ga$_9$ \cite{Shibayama,Wakui} and Ru$_7$B$_3$ \cite{Wen,Kase}. In the former class of materials, the electron correlations seem to play an important role in governing the superconducting properties. The latter class of materials is therefore more suitable for the study of the pure effects of inversion symmetry breaking.

In this Rapid Communication, we present the results of  NMR studies on the non-centrosymmetric superconductors Mg$_{9.3}$Ir$_{19}$B$_{16.7}$ ($T_c$=5.8 K) and Mg$_{10.5}$Ir$_{19}$B$_{17.1}$ ($T_c$=4.8 K). This material has a bcc crystal structure with the space group of  $I\bar{4}3m$.  There are two Mg sites, three Ir sites and two B sites. Among them, Ir(3)  site (24g-site), Mg(1) site  (8c-site) and the two B sites do not have inversion center \cite{Klimczuk,Xu}. In particular, Ir is a heavy element which may lead to a large spin-orbit coupling.  It has been reported that there is a wide range for stoichiometries; changing the stoichiometry only results in a small change in $T_c$ \cite{Klimczuk}. Specific heat and photoemission measurements suggested $s$-wave gap \cite{Mu,Li,Yokoya}, but there are also indications of exotic pairing from tunneling spectroscopy and penetration depth measurements \cite{Cava,penetration}. 
Our results of spin lattice relaxation rate ($1/T_1$) and the Knight shift indicate that the Cooper pairs are  predominantly in the spin-singlet state with an isotropic gap. 
However, Mg$_{10.5}$Ir$_{19}$B$_{17.1}$ is found to contain more defects and the Knight shift remains finite even in the zero-temperature limit. We interpret this result as  that the defect enhances the spin-orbit coupling and brings about more spin-triplet component. Our result suggests that properly introducing defects could provide a new route to exotic superconducting state.

\section{Experimental Procedures and Sample Characterization}
Two poly-crystal samples with different nominal composition were   prepared by the solid-state reaction method  with starting materials of Mg (99.8\% purity),  Ir (99.99\%) and B (99.7\%) \cite{Li}. The appropriate compositions of the starting materials powders were mixed and pressed into a pellet at a pressure of 1 GPa. Then the pellet was wrapped with Ta foil and sealed in an evacuated quartz tube, and  sintered at 600 $^o$C for 30 minutes and further at 950 $^o$C for 3 hours. The resultant pellet was well ground and pressed again, and  finally  was annealed at 950 $^o$C for 12 hours.
The inductively coupled plasma (ICP) analysis shows that sample 1 has a formula of Mg$_{9.3}$Ir$_{19}$B$_{16.7}$ and sample 2 is Mg$_{10.5}$Ir$_{19}$B$_{17.1}$. The uncertainty of the ICP analysis for the composition  is about $\pm$0.1.  The samples were also characterized by transmission electron microscope (TEM) spectroscopy. The samples for TEM measurement were prepared by crushing the powders in ethanol, and the resultant suspensions were dispersed on a holey carbon-covered Cu grid. The TEM investigation was performed on a FEI Tecnai-F20 (200kV) TEM.

For NMR measurements, the samples were crushed into powder. 
$T_c$ at zero and a finite magnetic field ($H$) was determined by measuring the ac susceptibility using the in-situ NMR coil. Figure 1 shows the result for $H$=0. 
$T_c(H=0)$  for Mg$_{9.3}$Ir$_{19}$B$_{16.7}$ is 5.8 K, which is a higher than the previous report by Klimczuk {\it et al} for nominal composition Mg$_{12}$Ir$_{19}$B$_{19}$ ($T_c$=5 K) \cite{Klimczuk},  and   $T_c(H=0)$  for Mg$_{10.5}$Ir$_{19}$B$_{17.1}$ is 4.8 K. 
\begin{figure}
\begin{center}
\includegraphics[scale=0.5]{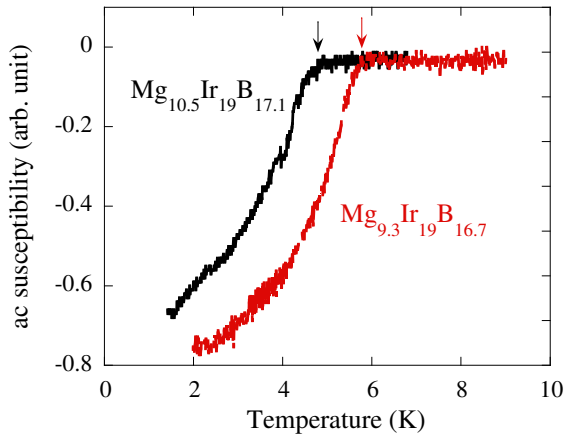}
\caption{(Color online) ac susceptibility measured using the in-situ NMR coil at zero magnetic field. The arrow indicates $T_c$ for each sample. }
\label{fig:1}
\end{center}
\end{figure}
A standard phase-coherent pulsed NMR spectrometer was used to collect data. 
In order to minimize the reduction in $T_c$ by $H$, the measurements were done at a low field of 0.44 T, at which  $T_c$ was reduced to 4.11 K and 3.45 K for the two samples, respectively.  The $^{11}$B NMR spectra were obtained by fast Fourier transform (FFT) of the spin echo, and were found to have a full width at half maximum (FWHM) of 4.6 kHz. The nuclear spin-lattice relaxation rate, $1/T_1$, was measured by using a single saturation pulse and by fitting the nuclear magnetization to a single exponential function since the quadrupole interaction is absent. 
 Measurements below 1.4 K were carried out in a $^3$He refrigerator. Efforts were made to avoid possible heating by the RF pulse,  such as using a small-amplitude   RF pulse.
%\begin{figure}
%\begin{center}
%\includegraphics[scale=0.4]{Fig1.eps}
%\caption{$^{11}$B NMR spectrum for Mg$_{9.3}$Ir$_{19}$B$_{16.7}$ and Mg$_{10.5}$Ir$_{19}$B$_{17.1}$  at $T$=4.2 K. }
%\label{fig:1}
%\end{center}
%\end{figure}
%
%Figure 1 shows the  NMR spectra for the two samples. The full width at the half maxima is 4.61 kHz and 4.85 kHz for Mg$_{9.3}$Ir$_{19}$B$_{16.7}$ ($T_c$=5.8 K) and Mg$_{10.5}$Ir$_{19}$B$_{17.1}$ ($T_c$=4.8 K), respectively. 
%

\section{Results and discussions }

 Figure 2 shows the temperature dependence of $1/T_1$  for the two samples. As can be  seen clearly in the figure, $1/T_1$   is enhanced just below $T_c$ over its normal-state value, forming a so-called coherence peak (Hebel-Slichter peak), which  is a hallmark of an isotropic superconducting gap.  Figure 3 shows  $1/T_1$ normalized by its value at $T_c$ against the reduced temperature, which compares the height of Hebel-Slichter peak of the two samples. 
\begin{figure}
\begin{center}
\includegraphics[scale=0.5]{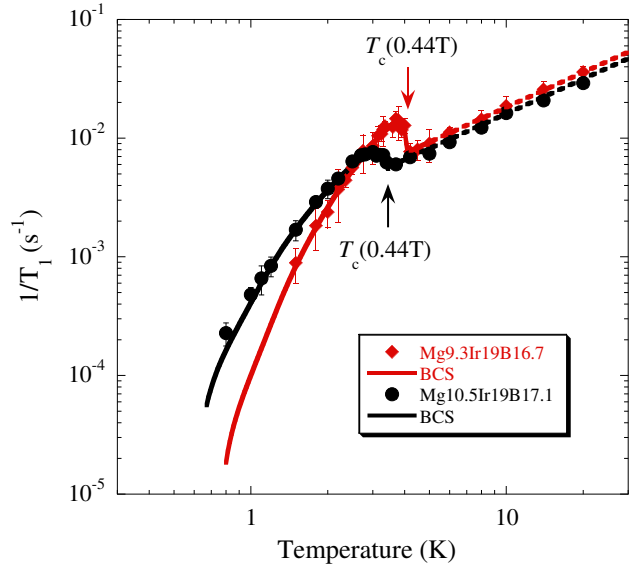}
\caption{(Color online) Temperature dependence of the $^{11}$B spin-lattice relaxation rate, $1/T_1$, in Mg$_{9.3}$Ir$_{19}$B$_{16.7}$ and Mg$_{10.5}$Ir$_{19}$B$_{17.1}$. The arrows indicate the superconducting transition temperature $T_c$ under the magnetic field of 0.44 T. The curves below $T_c$ are fits to the BCS theory with  2$\Delta_0=3.0k_BT_c$ (high-$T_c$ sample) and 2.2$k_BT_c$ (low -$T_c$ sample), respectively. The  broken lines above $T_c$ indicate the $1/T_1 \propto  T$ relation. }
\label{fig:2}
\end{center}
\end{figure}
 The $1/T_{1S}$ in the superconducting state is expressed as
%\begin{eqnarray}
$\frac{T_{1N}}{T_{1S}} = \frac{2}{k_BT}\int\int (1+\frac{\Delta^2}{EE'})N_s(E)N_s(E')f(E)[1-f(E')]\delta(E-E')dEdE'$
%\end{eqnarray}
where $1/T_{1N}$ is the relaxation rate in the normal state, $N_s(E)$ is the superconducting density of states (DOS),  $f(E)$ is the Fermi distribution function and $C =1+\frac{\Delta^2}{EE'}$   is  the "coherence factor". 
   Following Hebel \cite{Hebel}, we convolute  $N_{s}(E)$ with a broadening function $B(E)$ which is  approximated with a rectangular function centered at $E$ with a height of $1/2\delta$. The solid curves below $T_c$ shown in Fig. 2 and 3 are  calculations with 2$\Delta(0)=3.0k_BT_c$ and $r\equiv\Delta(0)/\delta$=5 for Mg$_{9.3}$Ir$_{19}$B$_{16.7}$, and 2$\Delta(0)=2.2k_BT_c$ and $r$=3 for Mg$_{10.5}$Ir$_{19}$B$_{17.1}$.  The smaller $\Delta(0)$ than the BCS value is probably due to the applied field. In Li$_2$Pd$_3$B, a smaller 2$\Delta(0)=2.2k_BT_c$ at a field of 1.46 T \cite{Nishiyama1} recovers to 3.0$k_BT_c$  at a smaller field of 0.44 T \cite{Kandatsu}.
\begin{figure}
\begin{center}
\includegraphics[scale=0.4]{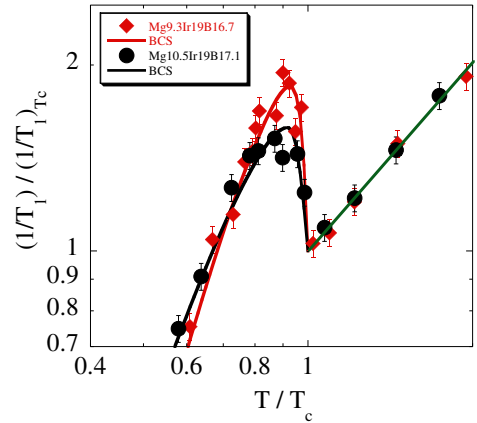}
\caption{(Color online) Normalized $1/T_1$   against the reduced temperature. The straight line above $T_c$ indicates the $1/T_1 \propto  T$ relation. }
\label{fig:3}
\end{center}
\end{figure}
\begin{figure}
\begin{center}
\includegraphics[scale=0.45]{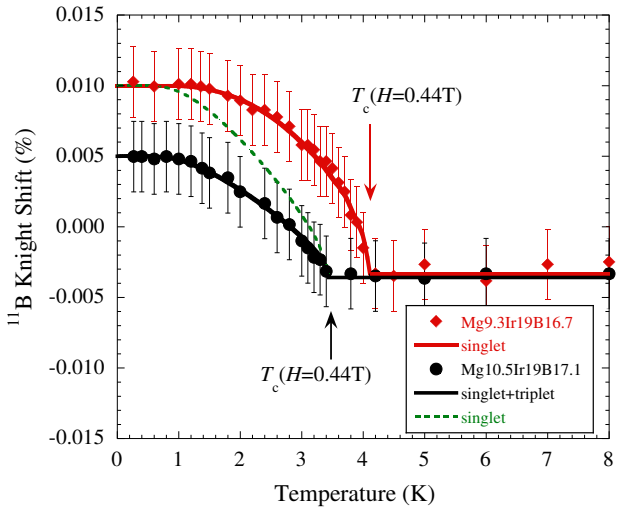}
\caption{(Color online) The $T$ dependence of the Knight shift for the two samples. The solid curve below $T_c$ for Mg$_{9.3}$Ir$_{19}$B$_{16.7}$ and the broken curve for  Mg$_{10.5}$Ir$_{19}$B$_{17.1}$ are calculations assuming purely singlet pairing. The solid curve for Mg$_{10.5}$Ir$_{19}$B$_{17.1}$ is a fit assuming mixing triplet and singlet pairings (see text for detail).}
\label{fig:4}
\end{center}
\end{figure}

Figure 4 shows the temperature dependence of the $^{11}$B Knight shift. Above $T_c$, the shift is temperature independent, while it changes below $T_c$. The observed Knight shift ($K_{obs}$) is composed of the spin part ($K_s$) and the orbital part ($K_{orb}$), $K_{obs}$=$K_s$+$K_{orb}$.    $K_{orb}$ is $T$ independent, and  $K_s$ is proportional to  $\chi_s$,   
$K_{s} = A_{hf}\chi_s$,            
where $A_{hf}$ is the hyperfine coupling between the nuclear and electron spins. In both samples, the shift increases below $T_c$.  
This indicates the decrease of $\chi_s$  in the superconducting state, since the hyperfine coupling constant is negative as seen in  Li$_2$Pd$_3$B \cite{Nishiyama1}. Thus the spin pairing in Mg-Ir-B systems is in the spin-singlet state. This is quite different from the case of Li$_2$Pt$_3$B \cite{Nishiyama}, although Ir and Pt are located next to each other in the periodic table. The difference is probably due to the fact that only 12/19  of Ir atoms sits in the non-centrosymmetric position. The solid curve below $T_c$ for Mg$_{9.3}$Ir$_{19}$B$_{16.7}$ and the broken curve for  Mg$_{10.5}$Ir$_{19}$B$_{17.1}$ in Fig. 4 are calculations assuming purely singlet pairing,
%\begin{eqnarray}
$\chi_s = -4\mu_B^2 \int N_s(E)\frac{\partial f(E)}{\partial E}dE$, 
%\end{eqnarray} 
with the same gap parameter obtained from $T_1$ fitting. 
In performing the fitting, $K_{orb}$=0.010\%  is assumed. 
It is a reasonable assumption that $K_{orb}$ does not depend on the composition.%  If one takes $K_{orb}$=0.10\%, the shift for Mg$_{10.5}$Ir$_{19}$B$_{17.1}$ should then follow the dotted curve as shown in Fig. 3. 
The experimental results thus indicates that there remains a finite spin susceptibility at $T$=0 for Mg$_{10.5}$Ir$_{19}$B$_{17.1}$. 

What is the origin of the finite spin susceptibility at $T$=0? We argue that defect or disorder is responsible for the finite spin susceptibility. Given that Mg$_{9.3}$Ir$_{19}$B$_{16.7}$ has a higher $T_c$, it can be assumed that this composition is close to the optimal stoichiometry. Then, the sample Mg$_{10.5}$Ir$_{19}$B$_{17.1}$ can be viewed as Ir deficient. 
%In fact, this sample has a wider NMR spectrum than Mg$_{9.3}$Ir$_{19}$B$_{16.7}$. 
TEM image  supports that  Mg$_{10.5}$Ir$_{19}$B$_{17.1}$ has more  defects. 

Figures 5(a) and 5(b) show respectively the electron diffraction patterns taken along [110], [001] zone-axis directions of the Mg$_{9.3}$Ir$_{19}$B$_{16.7}$ sample. All the diffraction spots in these patterns can be well indexed using the expected cubic unit cell with lattice parameters of $a$=10.57 $\AA$ (space group of $I\bar{4}3m$).   
%these results are consistent with the TEM and X-ray data as reported by T. Klimczuk et al \cite{Klimczuk}
%
In contrast, the electron diffraction patterns of Mg$_{10.5}$Ir$_{19}$B$_{17.1}$ always contain additional weak reflection spots following each fundamental spot, suggesting that this sample contains a rich variety of defect structure. Further HRTEM (high-resolution TEM) study suggests that these additional reflection spots are caused by Moire fringes and occurrence of local structural distortion in the sample. Figure 5(c) gives a typical electron diffraction pattern taken along [110] zone axis direction of Mg$_{10.5}$Ir$_{19}$B$_{17.1}$. Figure 5(d) shows a corresponding HRTEM image. The inset of Fig. 5(d) is the Fourier spectrum obtained by FFT, in which one of the extra spots is indicated by the arrow. Further careful FFT analysis indicates that the extra spots arise from the area marked as "B", which contains defect structure.

\begin{figure}
\begin{center}
\includegraphics[scale=0.3]{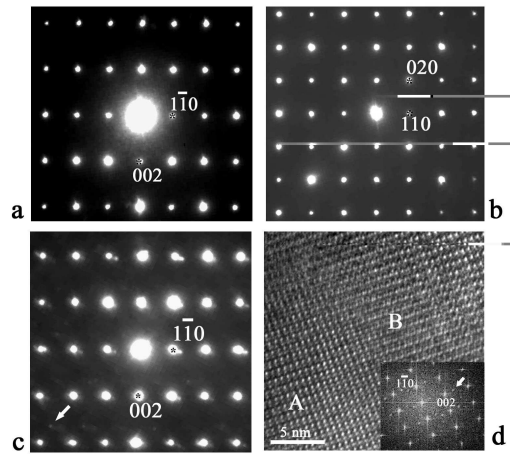}
\caption{(a) and (b): Electron diffraction patterns taken along [110], [001] zone-axis directions of Mg$_{9.3}$Ir$_{19}$B$_{16.7}$. (c): Electron diffraction pattern and (d) HRTEM image taken along [110] zone axis direction of Mg$_{10.5}$Ir$_{19}$B$_{17.1}$. The inset of (d) is the corresponding FFT pattern; one of the extra spots is indicted by the arrow.}
\label{fig:5}
\end{center}
\end{figure}
Now, in the presence of defect/disorder, there are two mechanisms that can give rise to the finite spin susceptibility. One is spin-reversal scattering by the impurity/disorder as pointed out by Anderson \cite{Anderson}. The other is mixing of the spin-triplet component due to enhanced spin-orbit coupling caused by defects. For spin-triplet superconductivity, the spin susceptibility does not decrease below $T_c$ or changes little, depending on the magnetic field configuration with respect to the $d$-vector that characterizes the triplet pairing. 
In the former case,  the isotropic scattering would reduce the gap anisotropy and would lead to an enhancement of the Hebel-Slichter peak, as was evidenced in Zn-doped Al \cite{Masuda}. However, this is not seen experimentally. In fact, as can be seen in  Fig. 3, the peak height is smaller in the low-$T_c$ sample Mg$_{10.5}$Ir$_{19}$B$_{17.1}$. Also, it seems hard to attribute a difference of 1 K in $T_c$ to non-magnetic impurity/defect in an $s$-wave superconductor.

We propose that the latter scenario, namely, the intrinsic effect of the defect that enhances SOC is more likely.  
The SOC is described by the Hamiltonian,
\begin{eqnarray}
H_{SO}= \frac{\hbar^2}{4m^2c^2}[\vec{\nabla}V(r)\times\vec{k}]\vec{\sigma}
\end{eqnarray}
where $\vec{k}$ and $\vec{\sigma}$ are the electron momentum and Pauli spin operator, respectively, and $\vec{\nabla}V (r)$ is the electrical field. In addition to the broken inversion symmetry, a vacancy of Ir can also  increase  $\vec{\nabla}V(r)$.  In particular, vacancies occupying the original centrosymmetric Ir(1) and Ir(2) sites will result in inversion-symmetry breaking for these sites and enhance the SOC. 
%The magnitude of the SOC depends on the number of positive charges ($Z$) that comprise the nucleus.
% and the averaged radius of an electron orbit. 
%As a good approximation, it goes in proportion to $Z^2$, which is about 3 times larger for Pt than Pd. 
The SOC lifts the two-fold spin degeneracy of the electron bands. As a result, the spin-singlet and spin-triplet states are mixed \cite{Gorkov,Frigeri,Sa,Fr}. 
 The extent to which the triplet-state component is mixed depends on the strength of SOC \cite{Gorkov,Frigeri,Sa,Fr}.  We propose that the finite spin susceptibility in Mg$_{10.5}$Ir$_{19}$B$_{17.1}$ is due to such SOC that is enhanced by Ir vacancy. The solid curve in Fig. 4 for Mg$_{10.5}$Ir$_{19}$B$_{17.1}$ is a fit assuming mixing triplet and singlet pairings, with finite $K_s$=0.005\% due to  triplet component, and the other $K_s$ due to singlet component with 2$\Delta$(0)=2.2$k_BT_c$.

\section{Conclusions}

In conclusion, we have presented the results of extensive NMR measurements on non-centrosymmetric superconductors Mg$_{9.3}$Ir$_{19}$B$_{16.7}$ ($T_c$=5.8 K) and Mg$_{10.5}$Ir$_{19}$B$_{17.1}$ ($T_c$=4.8 K). The spin lattice relaxation rate shows a coherence peak just below $T_c$ and follows an exponential $T$-variation at low temperatures. The spin susceptibility measured by the Knight shift decreases below $T_c$. These results indicate that the Cooper pairs are  predominantly in the spin-singlet state with an isotropic gap. This is likely due to the fact that only 12/19  of Ir atoms sits in the non-centrosymmetric position. However, Mg$_{10.5}$Ir$_{19}$B$_{17.1}$ is found to have more defects and the spin susceptibility remains finite even in the zero-temperature limit. We propose   that the defect enhances the spin-orbit coupling and brings about more spin-triplet component. We note that this mechanism may provide an alternative route to exotic superconducting state.

\section*{ACKNOWLEDGEMENT}

The authors wish to thank N. Ikeda for the suggestion of performing TEM experiment.  This work was partly supported by  research grants from MEXT and   JSPS (No. 17072005 and No. 20244058), and NSFC of China.   

\vspace{.1cm}

* Author to whom correspondence should be addressed: zheng@psun.phys.okayama-u.ac.jp

\end{document}